\documentclass{article}
\usepackage{graphicx}
\usepackage{amsmath}
\usepackage{bm}
\usepackage{cite}
\usepackage{hyperref}
\usepackage[english]{babel}

\def\rmi{\mathrm{i}}
\def\rme{\mathrm{e}}
\def\rmd{\mathrm{d}}

\def\psii{\bm{\psi}}
\def\R{\bm{\mathcal{R}}}
\def\T{\mathrm{T}}
\def\Tr{\mathrm{Tr}}
\def\w{\omega}
\def\mytextwidth{0.63\textwidth}

\newcommand\phsintgrnd[1][z]{q(#1)}
\newcommand\predexp[1][z]{q(#1)^{-1/2}}
\newcommand\phsintgrl[3][z]{\int_{#2}^{#3} \phsintgrnd[#1] \rmd #1}

\title{On a new exact relation for the connection matrices in case of 
a linear second-order ODE with non-analytic coefficients}
\author{Anton Kutlin\textsuperscript{1,*}}

\date{}

\begin{document}

\maketitle

\begin{center}
    {\bf 1}
    Institute of Applied Physics of Russian Academy of Sciences, 
    46 Ulyanov str., 603950 Nizhny Novgorod, Russia
    \\
    * anton.kutlin@gmail.com
\end{center}

\begin{abstract}
We consider the phase-integral method applied to an arbitrary linear ordinary second-order differential 
equation with non-analytic coefficients. We propose a universal technique based on the Frobenius method 
which allows to obtain new exact relation between connection matrices associated with its general solution.
The technique allows the reader to write an exact algebraic equation for the Stokes constants provided 
the differential equation has at most one regular singular point in a finite area of the complex plane. 
We also propose a way to write approximate relations between Stokes constants in case of multiple regular 
singular points located far away from each other. The well-known Budden problem is solved with help of this 
technique as an illustration of its usage. To access the HTML version of the paper \& discuss it with the author, visit
\url{https://enabla.com/pub/607}.
\end{abstract}

\section{Introduction \label{sec:intro}}
Consider an arbitrary ordinary linear differential equation written in the form of 
a stationary one-dimensional Schr\"odinger equation:
\begin{eqnarray}
\frac{\rmd^2 y}{\rmd z^2} + Q(z)y = 0.   \label{eq:gen}
\end{eqnarray}
The function $Q(z)$ will be referred to as a potential. An approximate local solution 
of equation \eqref{eq:gen} can be obtained with use of the phase-integral approximation\cite{frbook}. 
Provided that 
\begin{eqnarray}
\varepsilon = q^{-3/2} \rmd^2 q^{-1/2}/\rmd z^2  + (Q - q^2)/q \ll 1,   \label{eq:cond}
\end{eqnarray}
in the most general form the solution of \eqref{eq:gen} can be approximated by
\begin{subequations}
\begin{eqnarray}
y = c_+y_+ + c_-y_-, \label{eq:gensol}
\\
y_\pm = \predexp \exp [\pm \rmi \w(z)], \label{eq:phsint}
\\
\w(z)=\phsintgrl[\xi]{(z_0)}{z}, \label{eq:phsintgrl}
\end{eqnarray}
\end{subequations}
where the explicit form of $q(z)$ depends on the particular type of the approximation.
The simplest and the best-known type is one of the WKBJ\cite{wkb1,wkb2,wkb3,wkbj}; 
it takes the form \eqref{eq:phsint} with $\phsintgrnd = \sqrt{Q(z)}$.

The function $\w(z)$ is the phase integral, and therefore we call $\phsintgrnd[\xi]$ 
the phase integrand. A meaning of the brackets in the lower limit of integration in \eqref{eq:phsintgrl}
is a bit tricky; such a notation was introduced by Fr\"oman and Fr\"oman\cite{frpaper} to make
phase integral look similar for all orders of approximation. In the lowest order and, particularly,
in case of the WKBJ approximation, this integral is just a usual integral from $z_0$ to $z$. 

Clearly, the inequality \eqref{eq:cond} is not valid in a vicinity of poles or zeros of 
the phase integrand. Such points will be referred to as singular points and the 
vicinity of a singularity will be referred to as the interaction area.

Any phase-integral type solution \eqref{eq:gensol} is a local, not global, solution of \eqref{eq:gen}. 
The method of phase integrals allows the construction of a globally defined 
asymptotic expression for the solution of a desired linear ordinary differential 
equation. The method was first proposed by A.Zwaan\cite{zwaan} in his dissertation in 1929. 
He suggested allowing the independent variable in the differential equation to take 
complex values and to study a behaviour of the asymptotic solution far away from any 
singularities. According to Stokes\cite{stokes}, for any given exact solution 
of \eqref{eq:gen} coefficients $c_+$ and $c_-$ in the approximate solution \eqref{eq:gensol} 
differ from one domain of the complex plane to another 
(Stokes phenomenon \cite{stokes,rwbook,heading,frbook}). Such abrupt 
changes happen on the so-called Stokes lines and have a form of a single-parameter 
linear transform\cite{heading}. The parameter associated with a particular Stokes line 
is called the Stokes constant. Knowing all Stokes constants associated with a particular 
equation make it possible to obtain a globally defined approximate solution 
of \eqref{eq:gen} (see \cite{heading,rwbook}). A standard technique\cite{frpaper} used to obtain
equations for the Stokes constants is based on the assumption of a single-valuedness of
an exact general solution. This assumption fails in case of equations with regular singular points\cite{cbbook}, thus
the only way to solve such equations is to use some approximation for the Stokes constants\cite{rwbook, ours}
(e.g. the approximation of isolated singularities\cite{rwbook}). In the present paper, we provide a universal
technique to write equations for the Stokes constants. In case of a differential equation with
a single regular singular point, our equation is exact.

The paper is organized as follows.
In section \ref{sec:frob} we derive an exact equation for the Stokes constants in case of the differential
equation with the only one regular singular point. 
In section \ref{sec:budden} our technique is used to solve the well-known Budden problem. Also this
section contains a comparison of our result with an exact one and the result obtained with use
of the approximation of isolated singularities.
In section \ref{sec:approx} the possibility of writing approximate equation in case of multiple 
regular singular points is discussed.
And, finally, in section \ref{sec:con} are the conclusions. 

\section{The exact equation for the Stokes constants in case of a single regular singular point \label{sec:frob}}
Consider equation \eqref{eq:gen} with a single regular singular point. Without loss of generality,
we can place the singularity to the origin of the complex plane. A general solution of such an equation 
can be written in a form of the Frobenius series\cite{cbbook}:
\begin{subequations}
\label{eq:frobform}
\begin{eqnarray}
y(z) = A y_1(z)+B y_2(z), \label{eq:fgensol}
\\
y_1(z) = z^{f_1}\sum_{n=0}^{\infty}{a_n z^n},
\quad
y_2(z) = z^{f_2}\sum_{n=0}^{\infty}{b_n z^n} + K \ln(z) y_1(z), \label{eq:fy}
\end{eqnarray}
\end{subequations}
where $A$ and $B$ are arbitrary constants depending on particular boundary/initial conditions,  
$f_{1,2}$, $a_n$, $b_n$ and $K$ are coefficients determined by direct substitution of this form into the 
differential equation, and $a_0=b_0=1$ by convention. Particularly, this substitution gives
an indicial equation for the Frobenius indexes $f_1$ and $f_2$:
\begin{eqnarray}
f(f-1)+Q_{-2}=0,   \label{eq:indicial}
\end{eqnarray}
where $Q_{-2}$ is a coefficient before the inverse square of the complex variable $z$ in the Laurent series
of the potential. Since $z=0$ is the only singular point of the equation \eqref{eq:gen}, the Taylor 
series from \eqref{eq:fy} have infinite radii of convergence and the
Frobenius form \eqref{eq:frobform} is valid for any point of the entire complex plane.

Now consider a solution \eqref{eq:fgensol} with $A=1$ and $B=0$, i.e. $y(z)=y_1(z)$. 
As it can be seen from a substitution \mbox{$z \rightarrow z \rme^{2 \rmi \pi}$}, 
such a solution is an eigenfunction of the $2\pi-$rotation operator about the origin; 
corresponding eigenvalue is equal to $\rme^{2 \rmi \pi f_1}$.  
Then, suppose that for some values of $z$ in the vicinity of the complex infinity 
the solution can be written asymptotically in the form \eqref{eq:gensol} as
\begin{eqnarray}
y_1(z) \sim a_+y_+(z) + a_-y_-(z).
\end{eqnarray}
Using Heading`s rules of analytic continuation\cite{heading, rwbook} and a single-valuedness 
of squared phase integrand, we can write that
\begin{eqnarray}
y_1(z \rme^{2 \rmi \pi}) \sim b_+y_+(z) + b_-y_-(z),
\end{eqnarray}
where coefficients $b_{\pm}$ are expressed in terms of the Stokes constants, phase integrals and 
the coefficients $a_{\pm}$. Taking into consideration a linearity of equation \eqref{eq:gen}, 
the connection between $a_{\pm}$ and $b_{\pm}$ can be written in a simple matrix form as
\begin{eqnarray}
\psii_b = \R \psii_a,
\label{eq:rotation}
\end{eqnarray}
where $\psii_a = [{a_+,a_-}]^{\T}$, $\psii_b = [{b_+,b_-}]^{\T}$ and 'T' denotes the transpose operation.
The matrix $\R$ represents the $2\pi-$rotation operator; it is known in terms of
the Stokes constants and corresponding phase integrals. 

According to the reasoning above we can conclude
that we know exactly one of the $2\pi-$rotation matrix`s eigenvalues---the eigenvalue is 
equal to $\rme^{2 \rmi \pi f_1}$. Actually, we know another eigenvalue too. 
As it can be inferred from, for example, the F-matrix method\cite{frbook}, $\det\R=1$ and the second eigenvalue is equal 
to $\rme^{-2 \rmi \pi f_1}$. Now we can finally write the desired exact equation for the Stokes constants:
\begin{eqnarray}
\Tr\R = \rme^{2 \rmi \pi f_1} + \rme^{-2 \rmi \pi f_1} = 2 \cos(2 \pi f_1).
\label{eq:main}
\end{eqnarray}
In contrast to any known approximation, this equation reflects an actual brunching structure 
of the general solution of equation \eqref{eq:gen}. 

We defined the $2\pi$-rotation operator with help of the Heading`s rules and Stokes constants. However, 
according to \eqref{eq:rotation}, an existence of $\psii_a$ and $\psii_b$ is enough to the definition of
the operator regardless how exactly the $\psii-$vectors can be found. Since in the considered case of a 
single regular singular point a representation of the $2\pi-$rotation operator through the Stokes constants 
is exact, the equation for the Stokes constants is exact either. 

It is worth mentioning that $K$ from equation \eqref{eq:fgensol} may or may not be equal to zero. In the
former case, one can notice that we can infer the second eigenvalue of $\R$ right from the
Frobenius form solution; it is equal to $\rme^{2 \rmi \pi f_2}$. Then one can propose to write two
separate equations using two different eigenvalues instead of a single equation \eqref{eq:main}.
Actually, these two equations will not be independent; it can be seen right from the
indicial equation \eqref{eq:indicial}. Indeed, an explicit form of these two eigenvalues is
\begin{eqnarray}
\lambda_{1,2} = - \exp(\pm \rmi \sqrt{1 - 4 Q_{-2}}),
\end{eqnarray}
whence follows $\det\R=1$ and all the reasoning that led to equation \eqref{eq:main}.

\section{Example: the Budden problem \label{sec:budden}}
Consider a differential equation described the standard problem of penetration and 
resonant absorption of an electromagnetic wave, analyzed by Budden:\cite{white-chen,budden}
\begin{eqnarray}
\frac{\rmd^2 y(z)}{\rmd z^2} + \left( 1 + \frac{c}{z} \right) y = 0,  
\label{eq:budden}
\end{eqnarray}
where $y(z)$ is a complex amplitude of an electromagnetic wave.
The Budden problem is a problem of determination of the absorption in equation \eqref{eq:budden}. 
It can be solved exactly with use of an integral representation of the solution \cite{rwbook}.
Here we will examine this problem using the phase-integral method and the technique presented above. 

Since the WKBJ approximation is the simplest one,
we will use it all throughout this section. According to equation \eqref{eq:phsint}, 
equation \eqref{eq:budden} has $y_+ \propto e^{iz}$ and $y_- \propto e^{-iz}$ as its WKBJ asymptotic. 

Boundary conditions of equation \eqref{eq:budden} corresponding to the Budden problem can be
formulated as a presence of incident wave from the large negative $z$ and absence of such 
a wave from the large positive $z$. In terms of WKBJ asymptotics the conditions take the form
\begin{eqnarray}
y(z) \sim y_+(z), \quad z \rightarrow +\infty.  
\label{eq:bbound}
\end{eqnarray}
We define the reflection (transmission) coefficient $R$ ($T$) as
a ratio of the complex amplitudes of the reflected (transmitted) and incident waves. 
An absorption coefficient can be defined now as 
\begin{eqnarray}
A = 1 - |R|^2 - |T|^2.  
\label{eq:absdef}
\end{eqnarray}
Our aim in this section is to find the absorption coefficient.

To calculate the absorption using the phase integral method, we must express 
reflection and transmission coefficients in terms of the Stokes constants. 
For this purpose we must analytically continue our solution from the large positive $z$,
where we know its asymptotic due to our boundary condition \eqref{eq:bbound}, to
the large negative $z$. The continuation must be implemented through the lower half of the
complex plane---it follows from the prescription of the absorption`s positiveness\cite{rwbook}.

\begin{figure}
\centering
\noindent
\includegraphics[width=\mytextwidth]{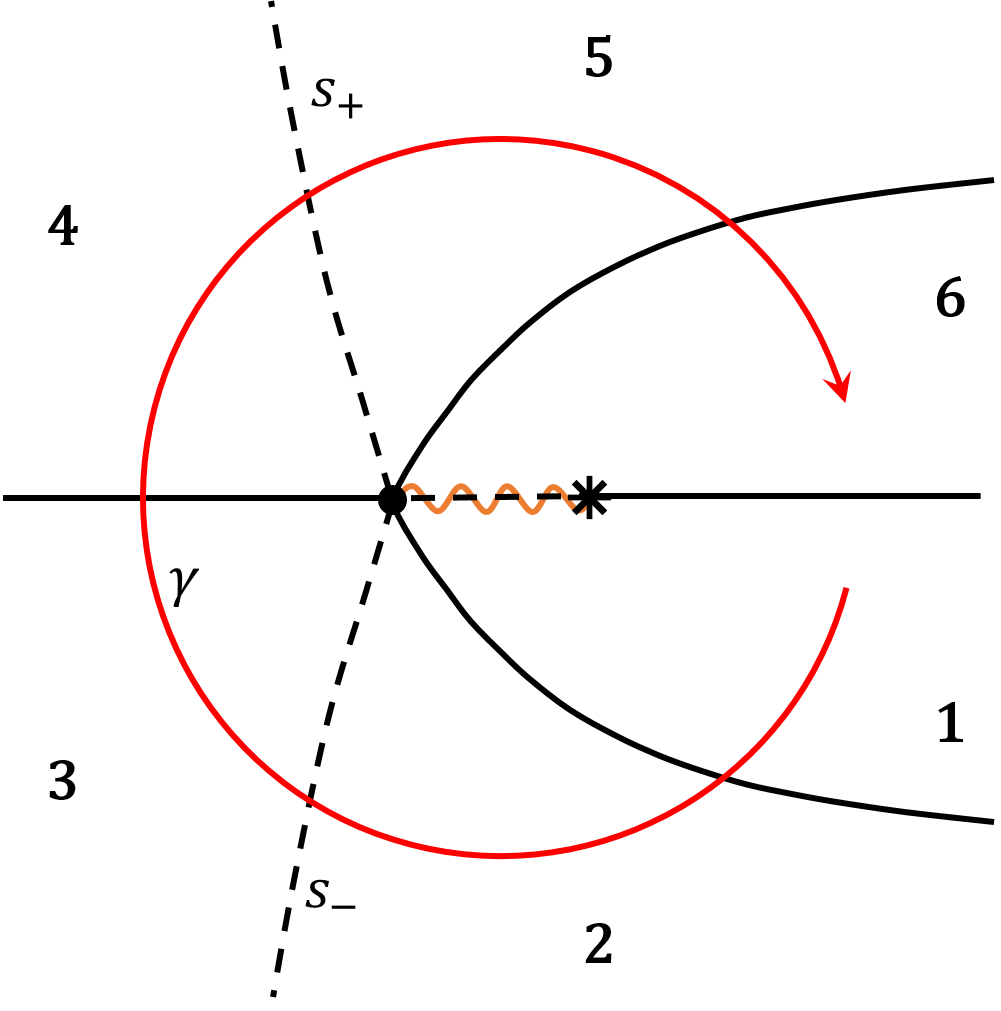}
\caption
{Stokes diagram for the Budden problem; Stokes lines are dashed.
A zero and a pole of the Budden potential are marked by a dot and a star correspondingly}
\label{fig:diagram}
\end{figure} 

The analytic continuation can be performed with use of the Heading`s rules. 
Following Heading\cite{heading}, define $(a,z)_{d,s}$ and $[a,b]$ as
\begin{eqnarray}
(a,z)_{d,s} = Q^{-1/4}\exp \left( \rmi \int_a^z \sqrt{Q} \rmd z \right),
\quad
\left[a,b\right] = \exp \left( \rmi \int_a^b \sqrt{Q} \rmd z \right), 
\end{eqnarray}
where subscripts $d$ and $s$ indicate correspondingly dominant and subdominant solutions 
in a given area of the complex plane. Let`s place a cut between $z=0$ and $z=-c$ 
along the real axis as shown on Fig.\ref{fig:diagram}. Begin from the large positive $z$ 
with $y=(0,z)$ we obtain:
\begin{equation}
\begin{split} 
&1.(0,z)_d=[0,-c](-c,z)_d 
\\
&2.[0,-c](-c,z)_d 
\\
&3.[0,-c](-c,z)_d - s_-[0,-c](z,-c)_s
\label{eq:ancont}
\end{split}
\end{equation}
where $s_-$ is a Stokes constant corresponding to the lower half of the complex plane.
The numeration in the analytic continuation \eqref{eq:ancont} corresponds to the
numeration of the complex plane`s areas in Fig.\ref{fig:diagram}.
All integrals here were evaluated below the cut and give $[0,-c]=\rme^{\frac{\pi c}{2}}$. As it
can be seen now from the continuation and the definition of the scattering characteristics,
\begin{equation}
y(z | Arg(z) = -\pi) \sim \rme^{\frac{\pi c}{2}}(-c,z) - s_-\rme^{\frac{\pi c}{2}}(z,-c),
\end{equation}
and
\begin{eqnarray}
R = -s_-, \quad T = e^{-\frac{\pi c}{2}}.
\label{eq:scattr}
\end{eqnarray}

To find $s_-$ and, thus, find $A$, we may use the branching structure preserving 
equation \eqref{eq:main} obtained in the previous section. The matrix $\R$ representing $2\pi-$rotation operator
can be found from the analytic continuation of the general solution of \eqref{eq:budden}
around the origin along the curve $\gamma$ (Fig.\ref{fig:diagram}). 
Starting with $y=a_+(0,z) + a_-(z,0)$ from large positive $z$, 
one can obtain that
\begin{equation}
\begin{split}
&1.\ a_+(0,z)_d+a_-(z,0)_s=a_+[0,-c](-c,z)_d + a_-[-c,0](z,-c)_s
\\
&2.\ a_+[0,-c](-c,z)_d + a_-[-c,0](z,-c)_s 
\\
&3.\ a_+[0,-c](-c,z)_d + (a_-[-c,0] - s_- a_+[0,-c])(z,-c)_s 
\\
&4.\ a_+[0,-c](-c,z)_s + (a_-[-c,0] - s_- a_+[0,-c])(z,-c)_d 
\\
&5.\ (a_-[-c,0] - s_- a_+[0,-c])(z,-c)_d 
\\&\qquad+ (a_+[0,-c] - s_+ (a_-[-c,0] - s_- a_+[0,-c]))(-c,z)_s
\\
&6.\ (a_-[-c,0] - s_- a_+[0,-c])(z,-c)_d 
\\&\qquad+ (a_+[0,-c] - s_+ (a_-[-c,0] - s_- a_+[0,-c]))(-c,z)_s.
\end{split}
\end{equation}
where $s_+$ is the Stokes constant corresponding to the upper half of the complex plane.
Finally, reconnecting from $z=-c$ to $z=0$ above the cut, we write that
\begin{equation}
\begin{split}
y(z | Arg(z) = -2\pi) \sim b_+(0,z) + b_-(z,0),
\\
b_+ = \rme^{c \pi} (1 + s_+s_-)a_+ - s_+a_-,
\\
b_- = s_- a_+ + \rme^{-c \pi} a_-.
\label{eq:R} 
\end{split}
\end{equation}
As long as the Budden potential has only a first order pole, equation \eqref{eq:main}
with $f_{1,2}=1,0$ takes the form $\Tr\R=2$ or, using the explicit form of $\R$ from Eq.~\eqref{eq:R},
\begin{eqnarray}
s_+s_- = - (1-\rme^{- \pi c})^2.
\label{eq:fbudres}
\end{eqnarray}

This nontrivial exact equation couldn't have been obtained from any other considerations. But we
still have only one equation and two unknowns. To proceed further we have to find one another
relation between the Stokes constants; and it could be done with use of the symmetry 
of equation \eqref{eq:budden}. The Budden potential is real on the real axis and
this fact allows us to write\cite{aksymm,frsymm} that
\begin{eqnarray}
s_+ = -s_-^\ast.
\label{eq:symmetry}
\end{eqnarray}
Now, considering equations \eqref{eq:scattr}, \eqref{eq:fbudres} and \eqref{eq:symmetry}, we can
finally obtain 
\begin{eqnarray}
A = e^{-\pi c}(1-e^{-\pi c}).
\label{eq:fabsorp}
\end{eqnarray}
The expression \eqref{eq:fabsorp} matches completely with the exact analytical result\cite{rwbook}.

\begin{figure}
\centering
\noindent
\includegraphics[width=\mytextwidth]{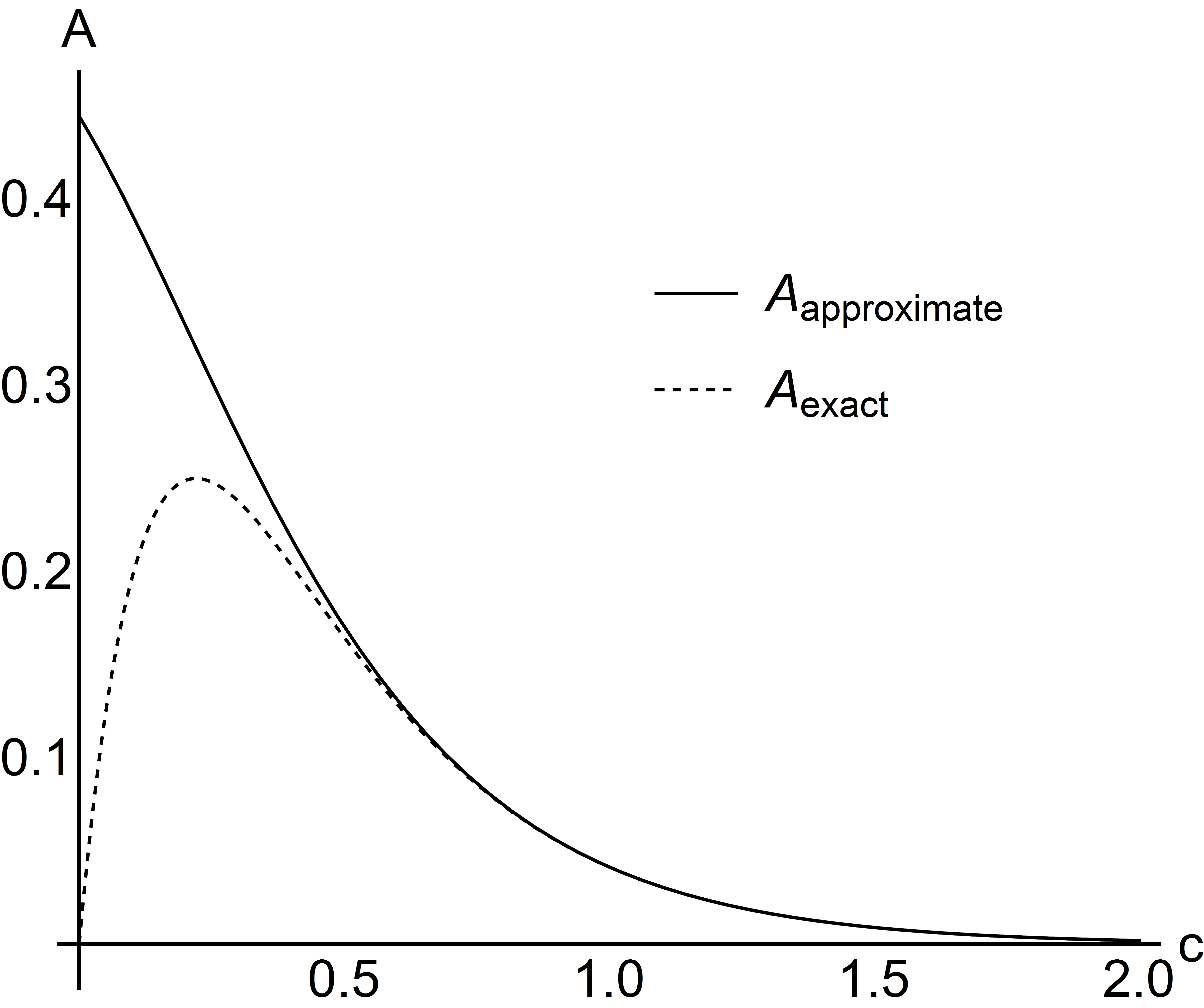}
\caption{Absorption efficiency in the Budden potential}
\label{fig:absorp}
\end{figure} 

It is worth mentioning that equations \eqref{eq:fbudres} and \eqref{eq:symmetry} do not allow us
to find a complex phase of the Stokes constants and, thus, a phase of the reflection coefficient.
However, this problem can be solved with use of the symmetry relations described in 
\cite{aksymm}. Completely analogous to the example of the Weber equation presented in
\cite{aksymm}, the Stokes constants for the Budden problem can be found exactly combining
the symmetry relations and the technique described in the present paper.

It can be useful to compare expression \eqref{eq:fabsorp} with an approximate result 
obtained with use of approximation of isolated singularities (Fig.\ref{fig:absorp}). According to the 
approximation\cite{rwbook}, reflection and transmission coefficients are
\begin{eqnarray}
R \approx 
-\rmi \frac{2 \rme^{\frac{\pi c}{2}} - \rme^{-\frac{\pi c}{2}}}{2 \rme^{\frac{\pi c}{2}} + \rme^{-\frac{\pi c}{2}}},
\quad
T \approx -\rmi \frac{2}{2 \rme^{\frac{\pi c}{2}} + \rme^{-\frac{\pi c}{2}}},
\end{eqnarray}
and for absorption
\begin{eqnarray}
A \approx \frac{4 \rme^{\pi c}}{(1 + 2 \rme^{\pi c})^2}.
\label{eq:aisabsorp}
\end{eqnarray}

Comparing expressions \eqref{eq:fabsorp} and \eqref{eq:aisabsorp}, we can see that 
they behave similarly when $c \geq 1$, 
but not when $c \ll 1$. The limit of $c\rightarrow 0$ is clearly wrong; in this limit $Q=1$
and there should be no reflection and absorption. The error is due to the use of the Stokes constants, 
which do not preserve the global structure of the solution with a branch point. 

\section{Approximate equation for the Stokes constants in case of multiple regular singular points \label{sec:approx}}

Now consider equation \eqref{eq:gen} with multiple regular singular points and place one of the points
to the origin of the complex plane. A solution of such an equation still can be written in
the Frobenius form \eqref{eq:frobform}, but the Taylor series in \eqref{eq:fy}
are necessarily converge only up to the nearest pole of the potential\cite{cbbook}. This fact
prevents us from writing any exact equation in the manner of Sec.\ref{sec:frob} because
we cannot infer any information about the $2\pi-$rotation operator. However, provided some
zeros of the potential grouped around the singular regular point forming a cluster and this cluster is
well-isolated from any other singularities, we can analytically continue our general solution
around this cluster and write an equation analogous to equation \eqref{eq:main}. Such an equation
is exact provided we use an exact representation of the analytic continuation operator, but
for the Stokes constants it is approximate since the representation of the operator
through the Stokes constants is approximate; the equation becomes more precise as the cluster
becomes more isolated. The precision of the equation can be inferred from the Fr\"omans estimates 
for the limiting form of the connection matrices\cite{frbook}; we claim that the approximate equation 
for the Stokes constants is applicable provided a phase-integral approximation itself is applicable 
for any point of the path chosen for the analytic continuation around the cluster.
 
\section{Conclusion \label{sec:con}}
The method of phase integrals is a beautiful and powerful method of a linear ordinary 
differential equation`s asymptotic analysis, but its range of applicability is highly 
restricted to relatively simple problems; more complicated problems need additional equations 
for the Stokes constants. Traditional equations for the Stokes constants, proposed by Fr\"omans\cite{frpaper},
are based on the assumption of a single-valuedness of a general solution. This assumption clearly fails
in case of a differential equation with regular singular points. We provide an alternative universal
technique based on the Frobenius method which allows to overcome this difficulty and to obtain the 
desired equation even in case of the regular singular point`s presence. 
Our technique let the reader write an exact algebraic equation for the Stokes constants in case
of a single regular singular point and an approximate equation when there are more than one such points.
The equation written in its general form without a specific representation of the operator of the analytic
continuation around regular singular point is always exact regardless how many poles the potential has. 

To conclude with, we have to mention that our method is formally applicable to equations without any regular
singular points; in this case, equation \eqref{eq:main} takes the form $\Tr \R = 2$. Certainly, this equation,
as well as two more, can be obtained from the traditional approach\cite{frbook}. However, we claim that this is the only equation
directly connected with the analytic continuation around the cluster of singularities; as it was shown,
for example, in \cite{aksymm}, two remaining equations follow from the symmetry relations.

\paragraph{Acknowledgments.}
This work was supported by RFBR (project No~15-02-07600).

\end{document}